%% file: 0paper.tex
\documentclass[acmsmall, screen, nonacm]{acmart}

\input{00packages.tex}

\AtBeginDocument{%
  \providecommand\BibTeX{{%
    \normalfont B\kern-0.5em{\scshape i\kern-0.25em b}\kern-0.8em\TeX}}}

\acmJournal{TSC}
\acmVolume{0}
\acmNumber{0}
\acmArticle{0}
\acmYear{2026}
\acmMonth{0}

\begin{document}


\title[Not My Truce: Personality Differences in AI-Mediated Workplace Negotiation]{Not My Truce: Personality Differences in AI-Mediated Workplace Negotiation}

\author{Veda Duddu}
\orcid{0009-0001-6443-6239}
\affiliation{%
 \institution{University of Illinois Urbana-Champaign}
 \city{Urbana}
 \state{IL}
 \country{USA}}
\email{vduddu2@illinois.edu}

\author{Jash Rajesh Parekh}
\orcid{0000-0003-3310-4634}
\affiliation{%
 \institution{University of Illinois Urbana-Champaign}
 \city{Urbana}
 \state{IL}
 \country{USA}}
\email{jashrp2@illinois.edu}

\author{Andy Mao}
\orcid{0009-0007-6060-9730}
\affiliation{%
 \institution{University of Illinois Urbana-Champaign}
 \city{Urbana}
 \state{IL}
 \country{USA}}
\email{hanqim2@illinois.edu}

\author{Hanyi Min}
\orcid{0000-0002-0095-8513}
\affiliation{%
 \institution{University of Illinois Urbana-Champaign}
 \city{Urbana}
 \state{IL}
 \country{USA}}
 \email{hanyimin@illinois.edu}

\author{Ziang Xiao}
\orcid{0000-0003-3368-0180}
\affiliation{%
 \institution{Johns Hopkins University}
 \city{Baltimore}
 \state{MD}
 \country{USA}}
 \email{ziang.xiao@jhu.edu}

\author{Vedant Das Swain}
\orcid{0000-0001-6871-3523}
\affiliation{%
 \institution{New York University}
 \city{New York}
 \state{NY}
 \country{USA}}
 \email{v.das.swain@nyu.edu}

\author{Koustuv Saha}
\orcid{0000-0002-8872-2934}
\affiliation{%
 \institution{University of Illinois Urbana-Champaign}
 \city{Urbana}
 \state{IL}
 \country{USA}}
\email{ksaha2@illinois.edu}

\renewcommand{\shortauthors}{Veda Duddu et al.}


\input{0abstract}

\ccsdesc[500]{Human-centered computing~Empirical studies in HCI}

\keywords{workplace, AI coach, personality traits, professional negotiations}

\maketitle


\input{1introduction} 
\input{2relatedwork} 
\input{3study}
\input{5RQ1}
\input{5RQ2}
\input{5RQ3}
\input{6discussion}
\input{7limitations}

\input{8conclusion}




\bibliographystyle{ACM-Reference-Format}
\bibliography{0paper}



\end{document}

\endinput

%% file: 00packages.tex
\usepackage{color, colortbl, xcolor}
\usepackage{url}
\usepackage{subcaption}
\usepackage{textcomp}
\usepackage{soul}
\usepackage{multirow}
\usepackage{enumitem}
\usepackage{mathtools}
\usepackage{siunitx}
\usepackage{array}
\usepackage{colortbl}
\usepackage{hhline}
\usepackage{wrapfig}
\usepackage{soul}

\usepackage{booktabs} 
\usepackage{array}
\usepackage{xcolor}

\newcommand*{\rowstyle}[1]{
  \gdef\@rowstyle{#1}%
  \@rowstyle\ignorespaces%
}

\newcolumntype{=}{
  >{\gdef\@rowstyle{}}%
}

\newcolumntype{+}{
  >{\@rowstyle}%
}

\usepackage{arydshln}
\definecolor{LightGray}{gray}{0.97}
\definecolor{linkColor}{RGB}{6,125,233}
\definecolor{green}{rgb}{0.0, 0.65, 0.31}
\definecolor{bleudefrance}{rgb}{0.19, 0.55, 0.91}
\definecolor{ceruleanblue}{rgb}{0.16, 0.32, 0.75}
\definecolor{grey}{HTML}{969696}
\definecolor{violet}{HTML}{756bb1}
\definecolor{dgrey}{HTML}{01665e}
\definecolor{lgrey}{HTML}{5ab4ac}
\definecolor{dgreen}{HTML}{005a32}
\definecolor{purple}{HTML}{ae017e}

\definecolor{editCol}{HTML}{000000}
\definecolor{maskCol}{HTML}{c51b7d}
\definecolor{lrColor}{HTML}{8856a7}
\definecolor{trColor}{HTML}{d01c8b}
\definecolor{ctColor}{HTML}{4dac26}
\definecolor{brickred}{HTML}{f03b20}
\definecolor{improveCol}{HTML}{253494}
\definecolor{worsenCol}{HTML}{d7191c}
\definecolor{DarkBlue}{HTML}{00008B}
\definecolor{mscolor}{HTML}{01665e}
\definecolor{nmscolor}{HTML}{bf812d}
\definecolor{lgreen}{HTML}{ccece6}
\definecolor{dolive}{HTML}{308014}

\definecolor{editCol}{HTML}{000000}
\definecolor{maskCol}{HTML}{c51b7d}
\definecolor{lrColor}{HTML}{8856a7}
\definecolor{trColor}{HTML}{d01c8b}
\definecolor{ctColor}{HTML}{4dac26}
\definecolor{brickred}{HTML}{f03b20}
\definecolor{improveCol}{HTML}{253494}
\definecolor{worsenCol}{HTML}{d7191c}
\definecolor{lgreen}{HTML}{e0f3db}
\definecolor{dpink}{HTML}{CD1076}
\definecolor{pink}{HTML}{FED2D2}
\definecolor{soothinggreen}{HTML}{4dac26}
\definecolor{darkred}{HTML}{8B0000}

\definecolor{dblue}{HTML}{104E8B}
\definecolor{violet}{HTML}{8A2BE2}
\definecolor{mscolor}{HTML}{01665e}
\definecolor{nmscolor}{HTML}{d8b365}
\definecolor{deepgrey}{HTML}{525252}
\definecolor{dslate}{HTML}{2F4F4F}
\definecolor{dolive}{HTML}{556B2F}
\definecolor{teal}{HTML}{388E8E}
\definecolor{mscolor}{HTML}{01665e}
\definecolor{nmscolor}{HTML}{d8b365}

\definecolor{aicolor}{HTML}{018571}
\definecolor{occolor}{HTML}{ff7799}

\definecolor{srcolor}{HTML}{e34a33}
\definecolor{smcolor}{HTML}{253494}
\definecolor{srsmcolor}{HTML}{7fcdbb}
\definecolor{bothcolor}{HTML}{fe9929}
\definecolor{onecolor}{HTML}{018571}
\definecolor{marroon}{HTML}{881c1c}

\colorlet{tablerowcolor4}{gray!50} 

\newcommand*{\textlabel}[2]{%
  \edef\@currentlabel{#1}
  \phantomsection
  #1\label{#2}
}
\usepackage{tcolorbox}

\colorlet{tableheadcolor}{gray!25} 
\colorlet{tablerowcolor}{gray!15} 
\colorlet{tablerowcolor2}{gray!45} 
\colorlet{tablerowcolor3}{gray!25} 

\newcommand{\rowcolmedium}{\rowcolor{tablerowcolor2}}
\newcommand{\rowcollight}{\rowcolor{LightGray}} %

\newif{\ifhidecomments}
  \hidecommentsfalse 
\ifhidecomments
    \newcommand{\veda}[1]{}
    \newcommand{\andy}[1]{}
    \newcommand{\jash}[1]{}
    \newcommand{\haylee}[1]{}
    \newcommand{\ziang}[1]{}
    \newcommand{\vedant}[1]{}
    \newcommand{\koustuv}[1]{}
\else
    \newcommand{\veda}[1]{\textbf{\small\sffamily{\textcolor{DarkBlue}{[#1 -- Veda]}}}}
    \newcommand{\andy}[1]{\textbf{\small\sffamily{\textcolor{dgreen}{[#1 -- Andy]}}}}
    \newcommand{\jash}[1]{\textbf{\small\sffamily{\textcolor{dolive}{[#1 -- Jash]}}}}
    \newcommand{\haylee}[1]{\textbf{\small\sffamily{\textcolor{violet}{[#1 -- Haylee]}}}}
    \newcommand{\ziang}[1]{\textbf{\small\sffamily{\textcolor{brickred}{[#1 -- Ziang]}}}}
    \newcommand{\vedant}[1]{\textbf{\small\sffamily{\textcolor{soothinggreen}{[#1 -- Vedant]}}}}
    \newcommand{\koustuv}[1]{\textbf{\small\sffamily{\textcolor{dpink}{[#1 -- Koustuv]}}}}
  \fi

\newcommand{\trc}{\texttt{Trucey}}

\newcommand{\cgpt}{\texttt{Control-AI}}
\newcommand{\hbk}{\texttt{Control-NoAI}}

\renewcommand{\textrightarrow}{$\rightarrow$}

\newcommand{\cze}{$\mathtt{C_0}$}
\newcommand{\cone}{$\mathtt{C_1}$}
\newcommand{\ctw}{$\mathtt{C_2}$}

\colorlet{tableheadcolor}{gray!25} 
\colorlet{tablerowcolor}{gray!5} 

\definecolor{neutralCol}{HTML}{dd1c77}
\definecolor{neutralGreen}{HTML}{31a354}
\definecolor{NewBlue}{HTML}{1879ba}
\definecolor{bleudefrance}{rgb}{0.19, 0.55, 0.91}  
\definecolor{AfTrColor}{HTML}{0868ac}  
\definecolor{BfTrColor}{HTML}{a8ddb5}  

\definecolor{AfCtColor}{HTML}{b10026}  
\definecolor{BfCtColor}{HTML}{fd8d3c}

\graphicspath{ {figures/} }

\newcommand{\para}[1]{\vspace{0.25em}\noindent\textbf{#1}~}

%% file: 0abstract.tex

\begin{abstract}


AI-driven conversational coaching is increasingly used to support workplace negotiation, yet prior work assumes uniform effectiveness across users. 
We challenge this assumption by examining how individual differences, particularly personality traits, moderate coaching outcomes. 
We conducted a between-subjects experiment (N=267) comparing theory-driven AI (\trc{}), general-purpose AI (\cgpt{}), and a traditional negotiation \hbk{}. 
Participants were clustered into three profiles---resilient, overcontrolled, and undercontrolled---based on the Big Five personality traits and ARC typology. 
Resilient workers achieved broad psychological gains primarily from the handbook, overcontrolled workers showed outcome-specific improvements with theory-driven AI, and undercontrolled workers exhibited minimal effects despite engaging with the frameworks. These patterns suggest personality as a predictor of readiness beyond stage-based tailoring: vulnerable users benefit from targeted rather than comprehensive interventions. 
The study advances understanding of personality-determined intervention prerequisites and highlights design implications for adaptive AI coaching systems that align support intensity with individual readiness, rather than assuming universal effectiveness.

\end{abstract}

%% file: 1introduction.tex
\section{Introduction}
Negotiating in the workplace---whether for promotions, workload boundaries, or push back against managerial decisions---is recognized to be difficult and emotionally fraught~\cite{Bradley_Campbell_2016,grandey2000emotional}. 
These conversations shape career advancement, compensation trajectories, and workplace satisfaction, yet many information workers avoid such negotiations entirely or enter them under-prepared, resulting in foregone opportunities such as stagnant wages~\cite{hart2024but}. 
Foundational negotiation literature emphasizes tactical and strategic processes: structuring offers, managing process, and communicating strategically~\cite{Baber_2022, nadler2003learning}. 
Yet, these skill-based frameworks inadequately account for power-laden and 
socio-organizational complexities inherent in real-world upward negotiation~\cite{morrison2000organizational, Cortina_Magley_2003}. 
Workers operate within hierarchical systems where they rationally anticipate potential backlash such as being negatively perceived within the workplace, damaging relationships, or jeopardizing career prospects~\cite{morrison2000organizational, Milliken_Morrison_Hewlin_2003}. 
The resulting psychological burdens---increased stress, eroded confidence---undercut negotiation outcomes even in individuals who possess tactical knowledge, often manifesting as suppressed expectations or premature withdrawal from bargaining~\cite{Brooks_Schweitzer_2011,sullivan2006negotiator}.

Importantly, these psychological barriers do not affect all workers uniformly. 
The perceived costs and benefits of negotiation---and the strategies individuals use to cope with stress---vary systematically across workers with different demographic backgrounds and individual differences, including personality attributes~\cite{parent2021multilevel}.
Prior organizational and HCI research shows that such individual differences shape how workers interpret workplace demands, regulate emotion, and engage with technological tools, even under similar structural conditions~\cite{das2019multisensor,mark2016neurotics,saha2019libra,das2019birds}. 
By extension, the benefits and burdens introduced by workplace technologies are unlikely to be evenly distributed. 
Tools designed to support negotiation may empower some workers while inadvertently increasing cognitive or emotional load for others, depending on how their design aligns with users' underlying dispositions and needs.

Within the context of workplace negotiation coaching, traditional methodologies, such as training programs, handbooks, or one-on-one mentorships---have been shown to improve critical workplace outcomes such as promotions and salary increases~\cite{Baber_2022,brett2016negotiation}. 
However, such support is often resource-intensive, episodic, and difficult to scale.
Recent advances in generative AI and large language models (LLMs) have made AI coaching newly accessible for workplace negotiations, offering on-demand, conversational guidance that was previously limited to one-on-one coaching. 
Emerging research has highlighted AI's potential for supporting negotiation rehearsal, emotional preparation, and reflective practice~\cite{dasswain2025ai,shaikh2024rehearsal,das2024teacher,mckendrick2023virtual}. 
Yet, little is known about whether these AI coaching tools are equally effective across workers---or whether individual differences shape how such tools are experienced, engaged with, and ultimately perceived as useful.

To address the above gap, in this paper, we examine how individual differences—particularly personality, operationalized as stable trait dimensions such as the Big Five~\cite{costa1992neo}—shape workers' experiences with AI-based negotiation coaching. 
Drawing on the ARC typology~\cite{costa2002replicability}, we assign participants to clusters we term \textit{resilient}, \textit{overcontrolled}, and \textit{undercontrolled}, reflecting adaptive, constrained, and high-neuroticism profiles, respectively.

Focusing on two AI-coaching approaches---a theory-driven coach grounded in ~\citeauthor{brett2016negotiation}'s negotiation framework (\trc{}) \cite{brett2016negotiation} and a general-purpose conversational AI coach (\cgpt{})---we investigate who benefits from which form of support, how workers linguistically engage with these systems, and how they adopt or resist negotiation frameworks during coaching interactions. 
By moving beyond average effects, we aim to examine the heterogeneity in perceived effectiveness and engagement, offering insight into when AI coaching supports psychological readiness at work and when it may introduce friction or misalignment.
In particular, our study is guided by the following research questions (RQs):


\begin{enumerate}
    \item [\textbf{RQ 1:}] How do personality traits shape the perceived effectiveness of AI negotiation coaching?
    \item [\textbf{RQ 2:}] How do personality traits explain users' linguistic engagement with AI negotiation coaching?
    \item [\textbf{RQ 3:}] How do personality traits associate with the differential adoption of negotiation frameworks during AI-mediated coaching interactions?
    
\end{enumerate}

To address these questions, we conducted a between-subjects experiment ($N = 267$) where participants were randomly assigned to one treatment condition --- \trc{} (theory-driven AI coaching grounded in Brett's negotiation framework) --- or one of two control conditions: \cgpt{} (general-purpose conversational AI coaching) serving as an active control, or a static negotiation guide (\hbk{}) serving as a passive control. All participants prepared for workplace negotiation scenarios (salary raises, promotions, or time-off requests) and completed pre- and post-task measures of psychological outcomes including self-efficacy, psychological empowerment, and negotiation preparedness. 
We identified three distinct personality clusters using $k$-means clustering on participants' Big Five personality traits, then examined how each cluster responded to different coaching modalities across psychological outcomes (RQ1), linguistic engagement patterns (RQ2), and adoption of negotiation frameworks (RQ3).

Our findings reveal that the effectiveness of AI negotiation coaching is not universal and is shaped by individuals' personalities. While our theory-driven AI, \trc{}, fostered strategic development across personality clusters, these effects were uneven and did not consistently translate into psychological growth. 
For ``resilient'' individuals (high extraversion, agreeableness, conscientiousness, and openness, and low neuroticism), the traditional \hbk{} produced the most substantial psychological gains in empowerment, meaning, and self-determination, suggesting that for those with strong internal resources, self-directed engagement preserves autonomy better than interactive AI. 
Conversely, for ``overcontrolled'' individuals (low openness and extraversion), \trc{} offered a crucial ``novelty bypass,'' providing a theory-driven strategy that significantly boosted self-efficacy over general-purpose AI, even if their external evaluations did not always reflect this internal benefit. 
Notably, the ``undercontrolled'' individuals (high neuroticism; low agreeableness and conscientiousness) showed minimal psychological or linguistic adaptation across any condition, indicating a fundamental mismatch with self-directed digital interventions. Linguistically, ``resilient'' users showed increased engagement on select dimensions (e.g., verbosity and accessibility) when interacting with \trc{}, though these patterns were not uniform across all linguistic measures.

These findings carry significant implications for the design and deployment of AI-powered coaching systems, challenging the notion of a one-size-fits-all approach. Our work underscores that effective AI coaching necessitates deep personalization, not just in content, but in considering whether individuals are psychologically positioned to receive and benefit from assistance. It further reveals that even when users are receptive, optimal support may require a multi-tool approach, integrating the structured, reviewable content of traditional resources with the interactive rehearsal capabilities of AI. Ultimately, this research motivates a shift towards hybrid designs that adaptively scaffold support, address individual psychological barriers, and integrate diverse modalities to truly enhance preparedness.


%% file: 2relatedwork.tex
\section{Related Work}
\subsection{Personality and Individual Differences in Negotiation}

A substantial body of work shows that personality traits shape how individuals approach negotiation and evaluative interpersonal situations ~\cite{brett2016negotiation, bandura1977self, sharma2013role}. Negotiation theory highlights how stable traits interact with situational cues, cognitive biases, and perceptions of risk to influence strategy selection and outcomes~\cite{brett2016negotiation}. In particular, self-efficacy has been shown to predict willingness to initiate negotiation, persist through discomfort, and engage with challenging conversations~\cite{bandura1977self, sharma2013role}.

Parallel work in HCI demonstrates that personality also shapes how users engage with conversational and mixed-initiative AI systems. Prior studies show that user preferences, engagement, and trust vary as a function of personality traits and system characteristics. For example,~\citeauthor{volkel2021eliciting} found that users responded differently to chatbot personalities, while ~\citeauthor{cai2022impacts} showed that conscientiousness predicted greater trust in mixed-initiative systems compared to user-initiative ones~\cite{volkel2021eliciting, cai2022impacts}. More recently, ~\citeauthor{kuhail2024assessing} demonstrated that personality congruence effects in advising systems were concentrated among extroverted users~\cite{kuhail2024assessing}.

Together, this literature establishes that personality systematically shapes both negotiation behavior and engagement with AI systems. However, existing work largely examines preferences, trust, or engagement in isolation. Few studies investigate whether personality traits predict who benefits from which type of AI-mediated support, particularly in socially complex tasks such as negotiation. Our work builds on this literature by examining how personality moderates the effectiveness of distinct AI coaching approaches rather than assuming uniform benefit.

\subsection{AI-Driven Technologies for Workplace Communication}

Human-computer interaction (HCI) and computer-supported cooperative work (CSCW) research has increasingly examined AI-driven technologies to support communication, productivity, and wellbeing in organizational contexts~\cite{nepal2025survey, dasswain2020social,saha2021job}. Prior work spans a wide range of applications, including email and task management~\cite{mark2016email, miura2025understanding}, emotion detection and affective sensing~\cite{kaur2022didn, murali2021affectivespotlight}, conversational training systems~\cite{wilhelm2025managers, khadpe2024discern}, AI-supported workplace writing~\cite{lu2024corporate, kadoma2024role}, and interventions targeting stress, wellbeing, and organizational culture~\cite{howe2022design, dasswain2020social}. For instance, ~\citeauthor{howe2022design} demonstrated that the timing and framing of digital stress-reduction interventions significantly influence their effectiveness for information workers.

At the same time, a complementary line of work highlights that AI-mediated workplace tools do not benefit all workers equally. Prior studies show that AI systems can both empower and constrain workers depending on context and design choices~\cite{lee2015working, das2023algorithmic}, while concerns around surveillance and monitoring remain salient~\cite{ajunwa2017limitless, roemmich2023emotion}. Researchers have also documented variation in perceived agency, inclusion, and control in AI-mediated communication~\cite{kadoma2024role}, cultural misalignment with AI writing tools~\cite{basoah2025should}, and resistance to algorithmic advice in professional decision-making~\cite{longoni2019resistance, erlei2022s}.

This body of work suggests that the value of AI for workplace communication is situated rather than universal, shaped by individual perceptions, trust, and social context~\cite{pal2026we}. However, much of this literature focuses on task-level outcomes or general perceptions of AI systems. Less is known about how individual differences, such as personality traits, systematically condition the effectiveness of AI support in high-stakes interpersonal workplace interactions like negotiation. Our work contributes to this literature by explicitly linking individual differences to differential outcomes in AI-supported negotiation coaching.

\subsection{AI Coaching and Rehearsal for Negotiation}

Recent advances in generative AI have enabled scalable, conversational coaching systems that support negotiation and difficult workplace conversations. Prior work demonstrates that AI-powered rehearsal environments allow users to practice conflict scenarios and receive feedback in low-stakes settings~\cite{shaikh2024rehearsal}. Related systems provide structured guidance for negotiation preparation and strategy selection~\cite{shea2024ace}, while other studies examine perceptions of authenticity, trust, and usefulness in AI-assisted managerial training~\cite{wilhelm2025managers}. Beyond negotiation, research on virtual rehearsal and emotional labor shows that AI support can reduce psychological burden and facilitate skill development in interpersonal work~\cite{mckendrick2023virtual, dasswain2025ai}. Applications have also extended to specialized domains such as humanitarian negotiation, where researchers emphasize the need for domain expertise alongside generative models~\cite{ma2025chatgpt}.

Despite these advances, existing work primarily evaluates whether AI coaching is effective on average. Most studies emphasize tactical skill acquisition or knowledge transfer, with limited attention to heterogeneity in user response~\cite{shea2024ace, ma2025chatgpt}. As a result, it remains unclear which users benefit from structured, theory-driven guidance versus more open-ended or supportive coaching approaches.
Our study addresses this gap by examining how personality traits moderate the effectiveness of AI negotiation coaching. We move beyond average treatment effects to analyze how individual differences shape perceived usefulness, linguistic engagement, and adoption of negotiation frameworks during AI-mediated coaching interactions.

%% file: 3study.tex
\section{Study and Methods}


This study examines how personality moderates the effectiveness of AI-based negotiation coaching. Participants were recruited as part of a randomized experiment comparing negotiation coaching approaches, which included 267 participants allocated across one treatment condition (\trc{}, $n = 134$) and two control conditions (\cgpt{}, $n = 66$; \hbk{}, $n = 67$). \cgpt{} served as an active control isolating \trc{}'s theoretical scaffolding, while \hbk{} served as a passive control isolating \trc{}'s interactive delivery. Sample size was determined a priori following \citet{dattalo2008determining}, with an anticipated effect size of $f^2 = 0.15$, desired power of 0.90, a conservative estimate of $k = 30$ predictors reflecting the full set of variables examined in the study considered across exploratory analyses, and $\alpha = .05$, yielding a minimum required sample size of 226. Our final sample of 267 participants exceeds this threshold, providing adequate power for the planned analyses.
For RQ1, we used the full sample of 267 participants to compare outcomes across all three conditions. For RQ2 and RQ3, we focused on the 200 participants in the two AI coaching conditions with complete personality and outcome data, comprising 134 participants in the \trc{} condition and 66 in the \cgpt{} condition.

Participants were recruited through Prolific, an online research platform, and were required to be at least 18 years old, proficient in English, and have U.S. workplace experience, including experience working under a supervisor and initiating negotiations through asynchronous communication. The sample reflected diverse demographic and professional backgrounds, with participants representing various work domains (engineering/technology, healthcare, finance, education), employment types (full-time, part-time), and compensation structures (salaried, hourly). 
Demographic distributions are summarized in \autoref{tab:demographic_experimental}.

\begin{table}[t!]
\centering
\sffamily
\footnotesize
   \caption{Demographic distribution of $N$=267 participants in the experimental study.}  
   \label{tab:demographic_experimental}
\begin{tabular}{lrrrr}
\textbf{} & \textbf{Overall} & \textbf{\trc{}} & \textbf{\cgpt{}} & \textbf{\hbk{}} \\
\toprule
\rowcolmedium \multicolumn{5}{l}{\textit{Sex}}\\
~Male & 141 & 71 & 33 & 37\\
~Female & 121 & 61 & 32 & 28\\
~Other/Prefer Not to Say & 5 & 2 & 1 & 2\\
\rowcolmedium \multicolumn{5}{l}{\textit{Race/Ethnicity}}\\
~White & 181 & 94 & 38 & 49\\
~Black or African American & 50 & 29 & 14 & 7\\
~Hispanic or Latino & 6 & 2 & 3 & 1\\
~Asian & 10 & 2 & 3 & 5\\
~Mixed/Other & 20 & 7 & 8 & 5\\
\rowcolmedium\multicolumn{5}{l}{\textit{Educational Level}}\\
~Associate Degree & 29 & 16 & 7 & 6\\
~Bachelors Degree & 114 & 50 & 29 & 35\\
~Graduate Degree & 80 & 44 & 20 & 16\\
~Other & 44 & 24 & 10 & 10\\
\rowcolmedium \multicolumn{5}{l}{\textit{Employment Type}}\\
~Employed full-time & 197 & 99 & 46 & 52\\
~Employed part-time & 43 & 21 & 10 & 12\\
~Other & 27 & 14 & 10 & 3\\
\rowcolmedium \multicolumn{5}{l}{\textit{Work Domain}}\\
~Administration/Operations & 10 & 5 & 2 & 3\\
~Creative/Design/Marketing & 15 & 5 & 5 & 5\\
~Education/Training & 21 & 14 & 4 & 3\\
~Engineering/Technology/Software & 46 & 24 & 12 & 10\\
~Finance/Accounting/Banking & 24 & 10 & 5 & 9\\
~Government/Public Service & 9 & 2 & 3 & 4\\
~Healthcare/Medical & 26 & 13 & 7 & 6\\
~Legal/Compliance & 4 & 1 & 3 & 0\\
~Management/Leadership & 16 & 8 & 4 & 4\\
~Sales/Business Development & 23 & 17 & 3 & 3\\
~Research/Analytics & 6 & 2 & 1 & 3\\
~Other/Mixed & 67 & 33 & 17 & 17\\
\rowcolmedium \multicolumn{5}{l}{\textit{Compensation Type}}\\
~Salaried (fixed annual amount) & 124 & 62 & 34 & 28\\
~Hourly wages & 81 & 41 & 21 & 19\\
~Other & 62 & 31 & 11 & 20\\
\rowcolmedium \multicolumn{5}{l}{\textit{Work Experience}}\\
~0-1 year & 4 & 1 & 3 & 0\\
~1-3 years & 16 & 9 & 2 & 5\\
~3-5 years & 22 & 10 & 9 & 3\\
~5-7 years & 29 & 14 & 8 & 7\\
~7 years or more & 195 & 100 & 44 & 51\\
~No response & 1 & 0 & 0 & 1\\
\bottomrule
\end{tabular}
\Description[table]{This table provides a detailed demographic breakdown of the 267 participants in the experimental study. It categorizes the participants based on several key characteristics, including Sex, Race/Ethnicity, Educational Level, Employment Type, Work Domain, Compensation Type, and Work Experience. The data is presented for the overall study sample as well as for each of the three intervention conditions: Trucey, ChatGPT, and the traditional Handbook. The table shows the distribution of participants across these demographic and professional categories, ensuring the study's sample reflected a diverse set of backgrounds.
}
\end{table}

\subsection{Designing an Interactive Prototype: \trc{}}
\begin{figure}[t]
        \centering
        \includegraphics[width=\columnwidth]{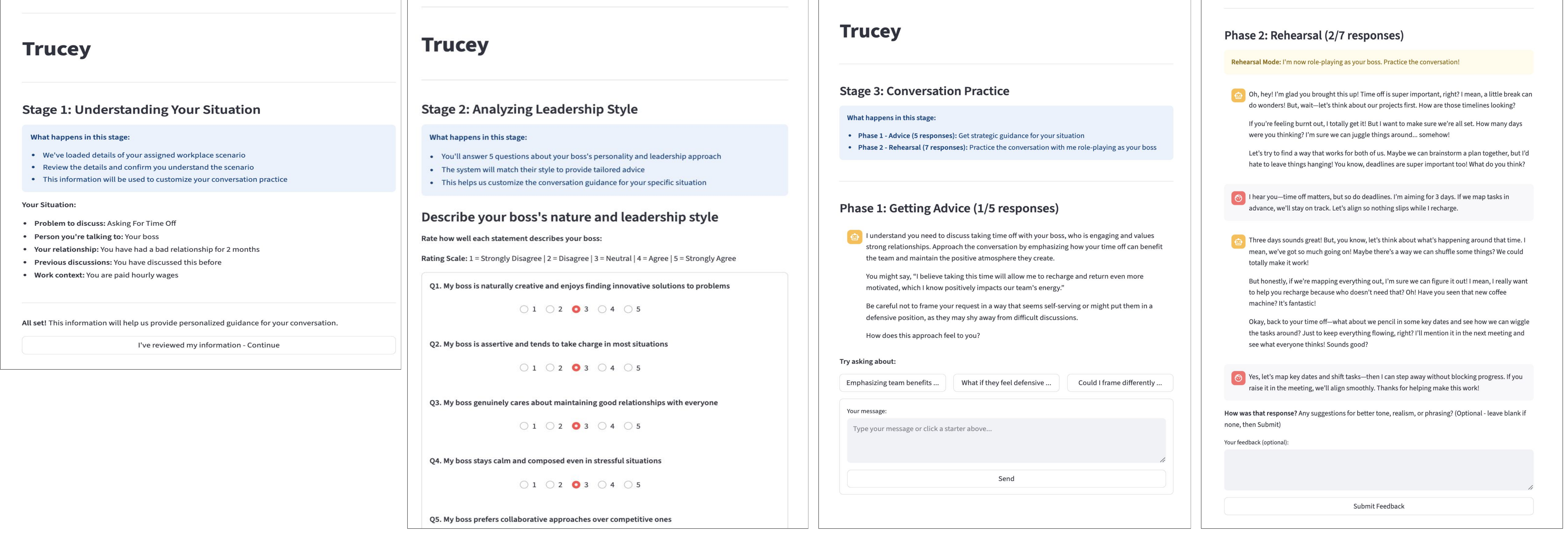}
    \Description[table]{This figure shows a series of screenshots from the Trucey prototype, demonstrating the sequential workflow a user experiences during a negotiation coaching session. The first screenshot, labeled "Stage 1: Understanding Your Situation," give the user their pre-assigned contextual details of a negotiation scenario, such as the problem to discuss, the person they're talking to, and their relationship with that person.
The second screenshot, "Stage 2: Analysing Leadership Style," shows a structured assessment where the user is asked to describe their boss’s nature and leadership style. This assessment, based on Big-Five personality traits, enables the system to tailor its strategic advice and simulation behavior to the specific supervisor's profile.
The third screenshot, "Stage 3: Conversation Practice," is where the core interaction occurs. It is divided into two phases: Phase 1: Getting Advice and Phase 2: Rehearsal. In the "Getting Advice" phase, Trucey provides strategic coaching through incremental dialogue, introducing negotiation concepts gradually to prevent user overwhelm. The "Rehearsal" phase is an interactive practice session where the AI embodies the supervisor's personality profile to create a realistic negotiation experience. 
This phase also includes an "Iterative Feedback Integration" step, which is shown at the bottom of the last screenshot. This feature allows users to provide direct feedback to the chatbot on its realism and tone. This feedback then helps to recalibrate the system for subsequent responses. The goal of this step is to make the coaching system more situationally responsive and to maintain a shared understanding between the user and the AI. The system is technically implemented using GPT-4.1 with a Python Streamlit interface and a SQL database to facilitate these personalized interactions.
}
    \caption{Example screenshots of \trc{} through various stages of user interactions.}
    \label{fig:screenshot}
\end{figure}

Our study centered on the design of an AI-powered interactive prototype for workplace negotiation, \trc{}. Its design was guided by two complementary theoretical frameworks from organizational science: 1)~\citeauthor{brett2016negotiation}'s negotiation framework emphasizes how contextual factors and power asymmetries shape the effectiveness of negotiation strategies~\cite{brett2016negotiation}, and 2)~\citeauthor{Bradley_Campbell_2016}'s framework for managing difficult conversation structures across the phases of preparation, in conversation management and post interaction reflection. 
These frameworks allowed us to implement \trc{} as a theory-driven AI coach providing strategic guidance along with conversational support. 
\trc{}'s prototype (\autoref{fig:screenshot}) was designed to provide contextually grounded coaching through a structured five-stage interaction workflow, described below:

\para{Stage 1: Scenario Assignment.} We assigned participants to one of three workplace negotiation scenarios: requesting a salary raise, asking for a promotion, or requesting time off. We systematically varied contextual details---such as the supervisor relationship, previous discussion history on the negotiation topic, and the context of their wages---to create diverse negotiation scenarios. 

\para{Stage 2: Personality Calibration.}  Participants used the BFI-10 scale to assess their real-life supervisor’s personality. This assessment characterized the supervisor across five dimensions~\cite{soto2017next}: extraversion, agreeableness, conscientiousness, neuroticism, and openness. By responding to items like ``My boss is reserved'' and ``My boss is generally trusting'' on a 5-point Likert scale, participants grounded the simulation in their actual workplace relationships.

\para{Stage 3: Theory-Guided Advice Generation.} \trc{} provides strategic coaching through a chat interface, delivering incremental advice across multiple turns. The coaching covers various negotiation pillars, including interests, alternatives, and legitimacy. Participants were allowed to ask questions and receive guidance tailored to their specific supervisor profile and situation, information that was collected/assigned in the previous stages. 

\para{Stage 4: Role-Based Rehearsal Simulation.} The AI system embodied the supervisor's personality profile to facilitate an interactive practice negotiation. The system responded as the supervisor would, reflecting their specific Big Five traits. For example, a highly agreeable supervisor collaborated more, while a highly neurotic supervisor expressed more resistance. Participants practiced their approach through this simulated conversation.

\para{Stage 5: Interactive Feedback Integration.} Throughout the advice and rehearsal phases, participants provided structured feedback every two turns. This allowed our participants to rate whether the interaction felt appropriately challenging, realistic or supportive. The system further used this feedback to calibrate its subsequent responses. 

\para{Technical Implementation.} We built the system using GPT-4o-mini via OpenAI's API, a Python Streamlit interface, and a SQL database to store configurations and log interactions. To match supervisor personalities, we transformed user responses into vector embeddings and applied cosine similarity to identify the most relevant profile from a pre-defined set. Each profile encoded specific communication patterns and negotiation behaviors. We employed few-shot prompting with examples spanning five difficulty levels, ranging from distributive tactics (e.g., ultimatums) to integrative strategies (e.g., collaborative framing).

For theory-driven prompting strategy, we applied few-shot learning with examples drawn from \citeauthor{brett2016negotiation}'s negotiation framework across five difficulty levels. These range from distributive tactics (e.g., ultimatums, direct demands) to integrative strategies (e.g., collaborative framing, interest alignment). 
This approach enables \trc{} to dynamically adjust advice and practice activities according to both user context and the selected supervisor profile. 
\subsection{Study Design and Procedure}
The study employed a between-subjects design with random assignment to either \trc{}, \cgpt{}, or \hbk{}. 
Participants completed a three-phase procedure: 

\para{Phase 1: Pre-Task Survey (5-7 minutes).}: Participants provided demographic information and completed baseline measures, including their own Big Five personality (BFI-10), occupational self-efficacy (OSS-6), psychological empowerment understanding (PEU), and negotiation preparedness (fear of and willingness to initiate negotiations).

\para{Phase 2: Coaching Interaction (15-20 minutes).} Participants entered one of the three randomly assigned conditions to interact with their designated coaching intervention. Although all participants received the same scenario and contextual details, the nature of the coaching interaction varied by condition (\autoref{sec: conditions}). 

\para{Phase 3: Post-Task Survey (5-7 minutes).} Participants repeated the psychological outcome measures (OSS-6, PEU, negotiation preparedness) to provide a basis for change assessment. They also rated system usability and the appropriateness of the intervention.


\subsection{Conditions}
\label{sec: conditions}
Participants were randomly assigned to one of three conditions:

\para{\trc{} (Theory-Driven AI-based negotiation coaching):} Participants used the \trc{} prototype and completed all five stages: situational calibration, personality calibration, structured advice, role-based rehearsal, and iterative feedback.

\para{\cgpt{} (General-Purpose AI-based negotiation coaching):} Participants used \cgpt{} through the same interface and model as \trc{}, but without the theoretical scaffolding, personality calibration, or structured mechanisms. Although participants received the same scenario details for grounding, the system provided no theoretically driven prompts or personality-based simulations. This design intends to isolate \trc{}’s theoretical mechanisms while maintaining an equivalent interface and model architecture.

\para{\hbk{} (Non-AI static negotiation coaching):} This is a static handbook-based resource that served as a control condition in isolating the theoretical content from interactive delivery. The participants received a written guide (approximately 1500 words) grounded in the same negotiation frameworks~\cite{brett2016negotiation, Bradley_Campbell_2016}. 
The handbook provided guidance on strategic framing principles, partnership language, ready-to-use scripts, preparation checklists and strategies for handling common challenges that may occur in negotiations. The guide addressed time-off requests, career advancement, and compensation discussions, though salary negotiation was framed within career advancement rather than as a standalone scenario, representing a difference in depth and framing relative to \trc{}'s dedicated salary raise scenario. Unlike \trc{}'s personalized coaching, the \hbk{} guidance was presented across all contexts. 

These conditions differ not only in interactivity but also in information structure, personalization, and scenario framing; therefore, comparisons should be interpreted as modality-level contrasts rather than isolating a single treatment factor.


\subsection{Data Collection}
We collected multiple forms of data during this study:

\para{Survey Data.} Through pre-and post-task surveys, we captured the participants' self-reported psychological outcomes, personality, demographics and system evaluations. All the surveys use validated instruments with Likert-scale responses. 

\para{Interaction Logs.} For the two AI conditions, our system logged all conversational turns, including user messages, system responses, feedback and timestamps. These logs allowed us to capture the complete interaction sequence allowing us to perform a complete linguistic analysis of the participant's engagement. 

\subsection{Measures}
We measured personality using the Big Five Inventory-10 (BFI-10)~\cite{rammstedt2007measuring}, occupational self-efficacy (OSS-6)~\cite{rigotti2008short}, psychological empowerment understanding (PEU)~\cite{spreitzer1995psychological}, and negotiation preparedness (fear and willingness to initiate). These measures used Likert scales (5-, 6-, or 7-point depending on the instrument). Personality was assessed pre-task only, while psychological outcomes were measured at both pre-task and post-task. Post-task surveys additionally included usability and system intervention appropriateness~\cite{finstad2010usability,weiner2017psychometric}.

We also extracted linguistic engagement metrics from conversation logs~\cite{dasswain2025ai, saha2025ai} and quantified framework adoption through BERT-based semantic similarity with Brett's negotiation elements~\cite{reimers2019sentence}.

\subsection{Analysis}
\subsubsection{Personality Clustering}
\begin{figure}[t]
\centering
\begin{subfigure}[b]{0.4\columnwidth}
    \centering
\includegraphics[width=\columnwidth]{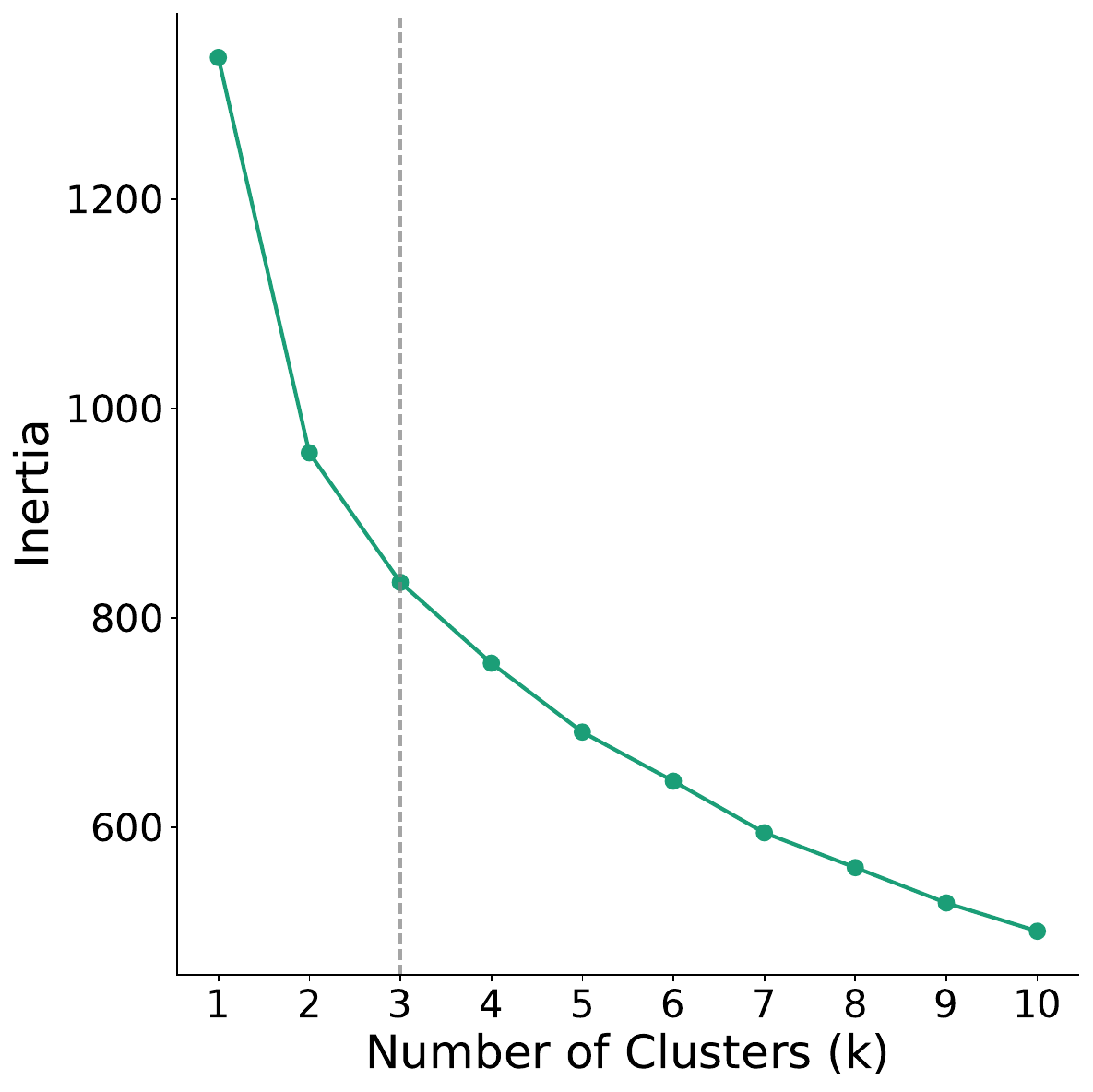}
    \caption{Elbow Plot}
    \label{fig:cluster_elbow}
    \end{subfigure}
\begin{subfigure}[b]{0.4\columnwidth}
    \centering
\includegraphics[width=\columnwidth]{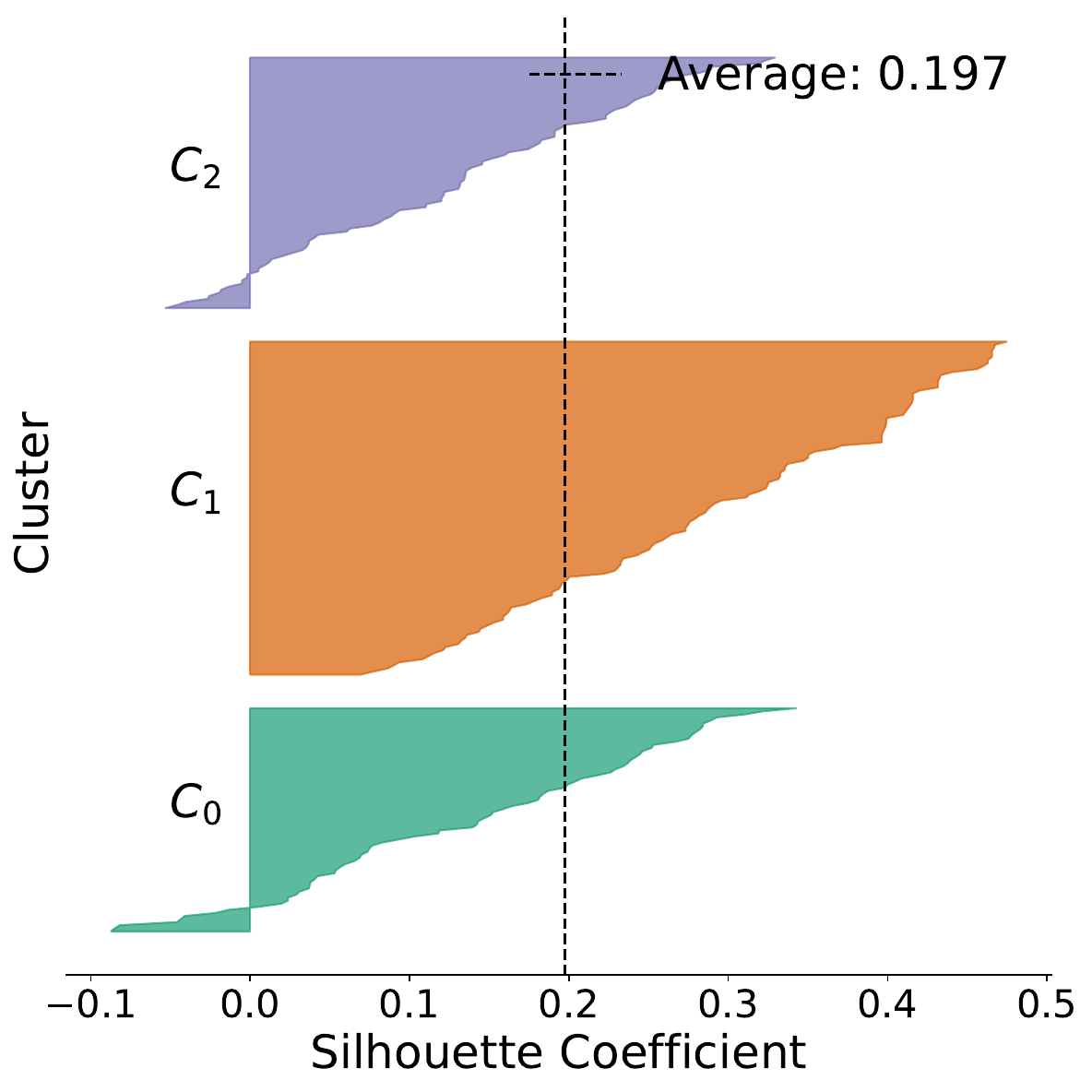}
    \caption{Silhouette Plot}
    \label{fig:cluster_silhouette}
    \end{subfigure}
    \begin{subfigure}[b]{0.4\columnwidth}
    \centering
\includegraphics[width=\columnwidth]{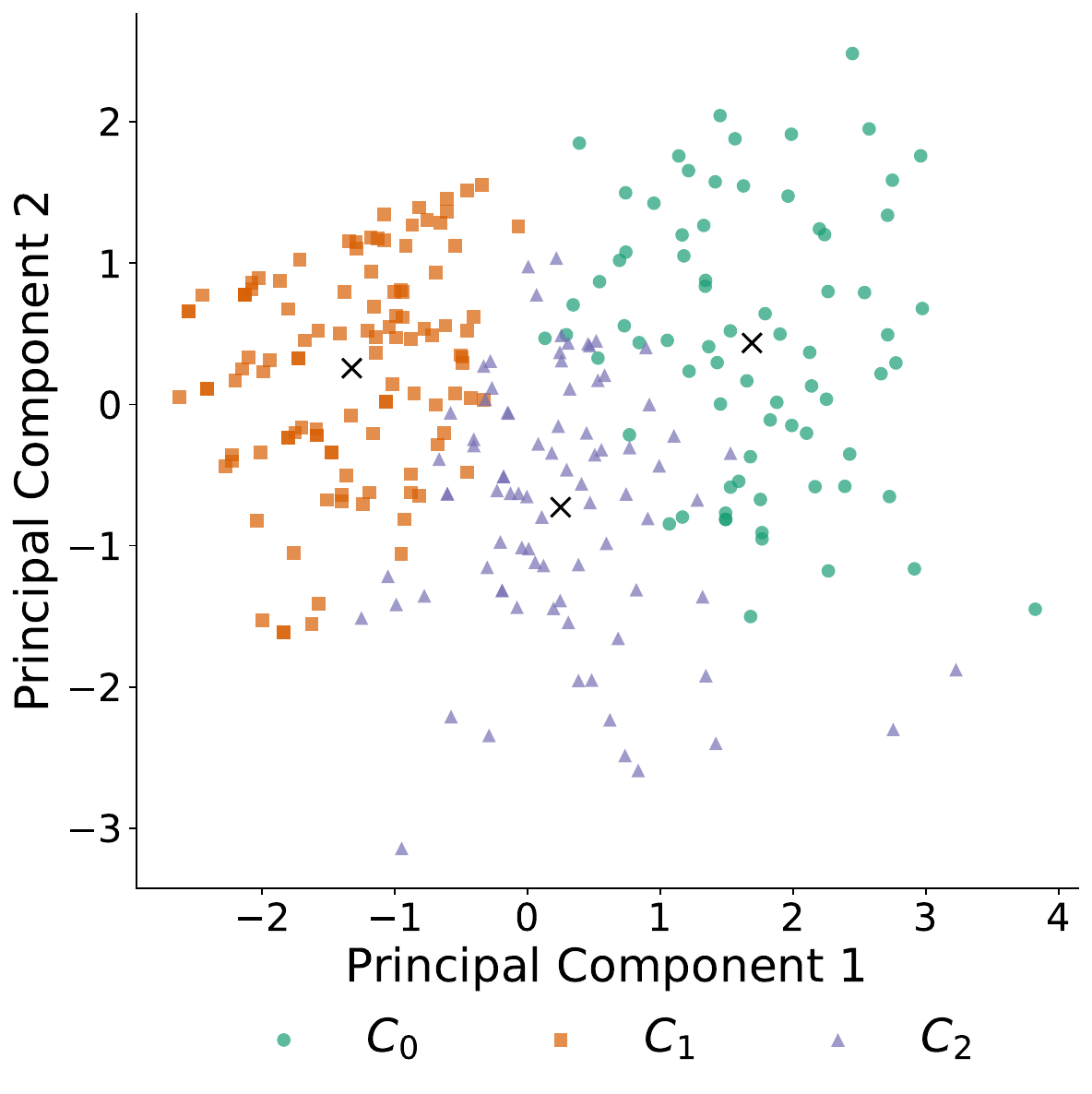}
    \caption{PCA distribution}
    \label{fig:cluster_pca}
    \end{subfigure}
        \begin{subfigure}[b]{0.4\columnwidth}
    \centering
\includegraphics[width=\columnwidth]{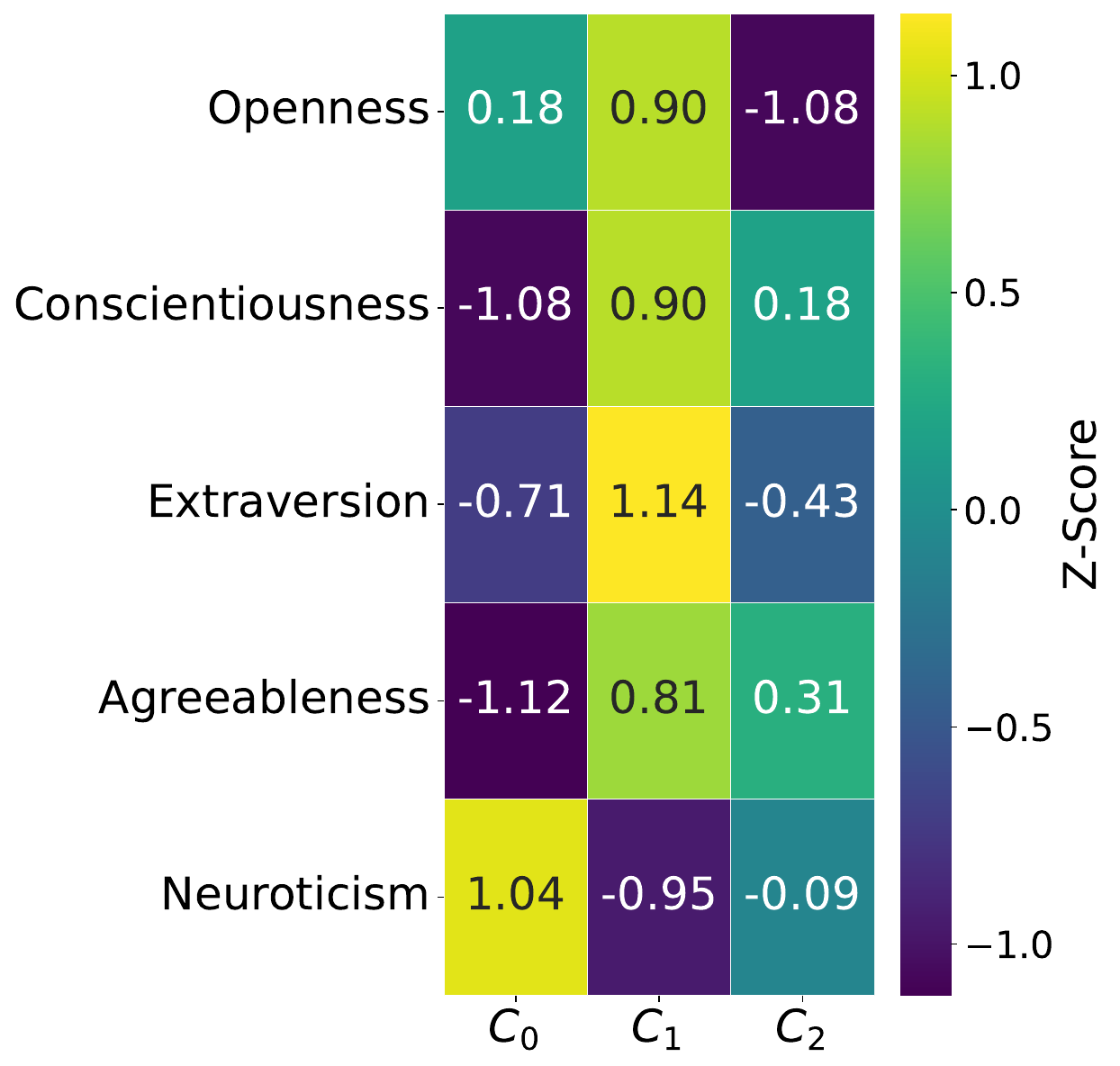}
    \caption{Heatmap}
    \label{fig:cluster_heatmap}
    \end{subfigure}
\caption{Figures describing the clustering and clusters: (a) Elbow method (b) Silhouette scores for determining the optimal number of personality clusters. (c) PCA-based distribution of the clusters across two PCA dimensions, and (d) Heatmap showing the distribution of the personality traits across the three clusters.}
\label{fig:cluster_validation}
\end{figure}

We applied $k$-means clustering to participants' $z$-standardized Big Five personality traits. We adopt a person-centered profile approach because our goal is to understand whether configurations of traits---rather than isolated trait effects---are associated with heterogeneous responses to AI-mediated coaching. Variable-centered analyses assume population homogeneity and can obscure meaningful subgroups, whereas person-centered approaches better capture how multiple characteristics act in concert to shape outcomes~\cite{bergman1997person, hofmans2020person, gillet2024multilevel}.
We evaluated cluster quality using the elbow method and silhouette scores (\autoref{fig:cluster_elbow} and \autoref{fig:cluster_silhouette}). We found an optimal solution at $k$=3 clusters. The silhouette score of 0.197 is consistent with prior Big Five clustering studies,  ~\cite{kerber2021personality} reported $k$-means silhouette scores of 0.079--0.084 on a sample of $N$=22,820 yet validated their solution using broader criteria --- and reflects the inherently fuzzy nature of ARC cluster structure, where types represent gradients of similarity to prototype profiles rather than discrete categories~\cite{chapman2011replicability}. \autoref{fig:cluster_heatmap} and \autoref{tab:cluster_centroids} show the distribution of the personality traits across the clusters. We draw on prior work~\cite{costa2002replicability, saha2024observer, dasswain2019multisensor} to describe and interpret these clusters.

\begin{table}[t]
\sffamily
\footnotesize
\centering
\caption{Cluster centroids ($z$-scored) for Big Five personality traits (N=267).}
\label{tab:cluster_centroids}
\begin{tabular}{ccccccc}
\textbf{Cluster} & \textbf{N} & \textbf{Extraversion} & \textbf{Agreeableness} & \textbf{Conscientiousness} & \textbf{Neuroticism} & \textbf{Openness} \\
\toprule
\cze{} & 74 & -0.52 & -1.04 & -0.96 & +0.94 & +0.06 \\
\cone{} & 110 & +0.61 & +0.58 & +0.61 & -0.66 & +0.56 \\
\ctw{} & 83 & -0.35 & +0.16 & +0.04 & +0.03 & -0.80 \\
\hline
\end{tabular}
\end{table}

\begin{itemize}
\item \textbf{Cluster \cze{}} \textit{(Undercontrolled)} comprises individuals with high neuroticism and low extraversion, agreeableness, and conscientiousness. 
This profile predisposes individuals to heightened anxiety in evaluative situations, diminishes cooperative engagement, and limits behavioral follow-through. These traits align with the ``undercontrolled'' configuration in the ARC taxonomy, which manifests as dysregulated responses to demanding interpersonal contexts~\cite{costa2002replicability}.

\item \textbf{Cluster \cone{}} \textit{(Resilient)} comprises individuals with high extraversion, agreeableness, conscientiousness, and openness, alongside low neuroticism.
These individuals exude positivity, maintain high levels of social engagement, and demonstrate strong psychological adjustment. They resist stress effectively and readily engage in autonomous self-regulation. The ARC taxonomy labels this profile ``resilient,'' a configuration that signals superior psychological adjustment~\cite{costa2002replicability}.

\item \textbf{Cluster \ctw{}} \textit{(Overcontrolled)} comprises individuals with low openness and low extraversion. 
These individuals prefer familiar approaches and structured environments over exploratory or novel engagement~\cite{costa1992neo}. This profile mirrors a documented variant of the ``Overcontrolled'' configuration, which features low levels of openness~\cite{costa2002replicability}.
\end{itemize}

\subsection{Privacy, Ethics, and Reflexivity}
Our study was reviewed and approved by the Institutional Review Board (IRB) at our university. To protect participants sharing sensitive workplace and interpersonal dynamics, we have implemented rigorous privacy and ethical safeguards. We collected the experimental dataset via Prolific and Qualtrics, ensuring anonymization at the point of collection. 

Our team brings together researchers from diverse gender, racial, and cultural backgrounds---including people of color and immigrants---spanning the fields of HCI, CSCW, computational social science, AI ethics, and I/O psychology. Our prior research on workplace dynamics and our collective lived experiences within diverse organizational settings directly inform our analysis. While we took the utmost care to faithfully synthesize participant viewpoints, we recognize that our interpretations remain situated within our disciplinary training and professional identities. We present these findings as an integration of participant voices and our own expert perspectives on the intersection of personality and technology.


%% file: 5RQ1.tex
\section{Results}
\subsection{RQ 1: How do personality traits shape the perceived effectiveness of AI-driven negotiation coaching?}
\label{sec:RQ1}

\begin{table}[t]
\sffamily
\footnotesize
\centering
\caption{Pooled cluster comparison across all conditions. Kruskal-Wallis tests examining whether personality clusters differed on outcome and perception measures when pooled across all conditions. $p$-values reported after Bonferroni correction. *** $p < .001$.}
\label{tab:pooled-cluster}
\begin{tabular}{lrrrr@{}l}
$\Delta$ \textbf{Metric} & \textbf{\cze{}}  & \textbf{\cone{}} & \textbf{\ctw{}} & \textbf{H-stat.} \\
\toprule
Organizational Self-Efficacy (OSS) & -0.041 & 0.012 & 0.011 & 0.17 \\
Psychological Empowerment (PEU) & 0.152 & 0.002 & 0.166 & 4.79 \\
\quad PEU: Competence & 0.099  & 0.027  & 0.036  & 0.60 \\
\quad PEU: Meaning & 0.223 & 0.018  & 0.199 & 1.47 \\
\quad PEU: Impact & 0.189 & 0.015  & 0.258  & 5.09 \\
\quad PEU: Self-Determination & 0.095 & -0.055 & 0.169 & 2.81 \\
\hdashline
\textit{Perception scores} \\
Usability Metric for User Experience (UMUX) & 69.14 & 82.65 & 76.40 & 26.72 & *** \\
Intervention Appropriateness Measure (IAM) & 3.56  & 4.26 & 4.01  & 31.75 & *** \\
\bottomrule
\end{tabular}
\end{table}

Our results demonstrate that AI coaching effectiveness is not a fixed property of the tool, but emerges from an interaction between the system's architecture and the user’s personality configuration. Before examining condition effects within clusters, we conducted a pooled Kruskal-Wallis test to compare the three clusters across all conditions (\autoref{tab:pooled-cluster}). 

We did not note any significant differences in any change score (all $p$ > 0.05), indicating that the three clusters did not differ in the overall magnitude of change when pooled across conditions. 
However, significant differences emerged in system perception: UMUX ($H$ = 26.72) and IAM ($H$ = 31.75). 
Across all conditions, with \cone{} (Resilient) rated the systems the highest while \cze{} (Undercontrolled) provided the lowest ratings. Within- and between- cluster analyses (\autoref{tab:within-cluster} and \autoref{tab:between-cluster-condition}) further reveal how specific traits dictate which intervention format translates into psychological gain.

\begin{table}[t]
\centering
\footnotesize
\sffamily
\caption{Within-cluster condition comparisons. Kruskal-Wallis $H$ tests examining whether conditions produced different outcomes within each personality cluster. Significant omnibus effects are bolded; 
* $p$ < 0.05, ** $p$ < 0.01, *** $p$ < 0.001}
\label{tab:within-cluster}
\begin{tabular}{lcrrrr@{}l}
$\Delta$ \textbf{Metric} &  \textbf{Cluster} &  \textbf{\trc{}} & \textbf{\cgpt{}}& \textbf{\hbk{}} & \textbf{H-stat.}\\
\toprule
Organizational Self-Efficacy (OSS) & \cze{} & -0.090 & -0.125 & 0.111 & 0.96 \\
& \cone{} & 0.047 &  0.037 & -0.090 & 0.96 \\
& \ctw{} & \textbf{0.200} & \textbf{-0.266}  & \textbf{-0.050} & \textbf{8.68}&\textbf{*} \\
\hdashline
Psychological Empowerment (PEU) & \cze{} & 0.087 & 0.174  & 0.248 & 1.93 \\
& \cone{} & \textbf{-0.086} & \textbf{-0.088} & \textbf{0.285} & \textbf{8.26} & \textbf{*} \\
& \ctw{} & 0.099 & 0.054 & 0.427 & 3.71 \\
\hdashline
\quad PEU: Competence & \cze{} & 0.072 & -0.125 & 0.317 & 3.49 \\
& \cone{} & -0.064 & -0.012 & 0.269 & 3.69 \\
& \ctw{} & -0.017 & 0.028 & 0.150 & 0.55 \\
\hdashline
\quad PEU: Meaning & \cze{} & 0.149 & 0.469 & 0.167 & 1.19 \\
& \cone{} & \textbf{-0.044} & \textbf{-0.204} & \textbf{0.385} & \textbf{6.87} & \textbf{*} \\
& \ctw{} & \textbf{0.112} & \textbf{-0.021} & \textbf{0.625} & \textbf{6.85} & \textbf{*} \\
\hdashline
\quad PEU: Impact & \cze{} & 0.072 & 0.167 & 0.413 & 4.54 \\
& \cone{} & 0.012 &  -0.062 &  0.103 & 0.28 \\
& \ctw{} & 0.250 &  0.074 &  0.483 &  3.26 \\
\hdashline
\quad PEU: Self-Determination & \cze{} & 0.054 & 0.188 &  0.095 &  0.29 \\
& \cone{} & \textbf{-0.246} & \textbf{-0.074} &  \textbf{0.385} &\textbf{9.69} & \textbf{**} \\
& \ctw{} & 0.050 & 0.133 &  0.450 & 2.24 \\
\hdashline
Usability Metric for User Experience (UMUX) & \cze{} & 63.514 & 72.917 & 76.190 & 5.21 \\
& \cone{} & 78.947 & 89.043 & 84.135 & 2.73 \\
& \ctw{} & 77.188 & 71.736 & 80.208 &  4.06 \\
\hdashline
Intervention Appropriateness Measure (IAM) & \cze{} & 3.399 &  3.766 &  3.702 & 2.53 \\
& \cone{} & 4.158 & 4.278 & 4.452 & 1.67 \\
& \ctw{} & 4.006 &  3.947 & 4.100 & 0.15 \\
\bottomrule
\end{tabular}
\end{table}

\para{Cluster \cze{} (Undercontrolled):} Undercontrolled individuals---those with high neuroticism and low agreeableness and conscientiousness---showed no significant condition effects on any psychological outcome (all $p$ > 0.09). These individuals are associated with dysregulated responses to demanding interpersonal stressors~\cite{costa2002replicability}. Within this cluster, change score means remained near zero across all conditions, suggesting that none of the interventions produced a consistent shift. High neuroticism likely heightens threat appraisal in these evaluative situations, while low agreeableness and conscientiousness limit cooperative engagement and behavioral follow-through. These compounding barriers likely create a state of fundamental unreadiness for self-directed digital interventions, where the lack of response reflects a mismatch between the intervention design and the user's baseline psychological needs.

\para{Cluster \cone{} (Resilient):} These are individuals who possess high extraversion, agreeableness, conscientiousness, and openness, and low neuroticism. For these well-adjusted individuals, the static \hbk{} produced significant positive gains in empowerment ($H = 8.26$, $p < .05$), meaning ($H = 6.87$, $p < .05$), and self-determination ($H = 9.69$, $p < .01$). Post-hoc comparisons confirmed that \hbk{} significantly exceeded \trc{} across all three dimensions: empowerment ($0.285$ vs. $-0.086$, $p_{\text{corr}} = .021$), meaning ($0.385$ vs. $-0.044$$p_{\text{corr}} = .044$), and self-determination ($0.385$ vs. $-0.246$, $p_{\text{corr}} = .006$). This increase in self-determination represents the largest single change in the dataset and emerged exclusively under the \hbk{} condition. High openness and conscientiousness likely drive a preference for self-directed engagement and informational support that preserves autonomy~\cite{ryan2000self}. This indicates that for individuals with high internal resources, interactive coaching functions as an autonomy constraint, whereas reference-based content allows them to leverage their existing self-regulation skills.

\para{Cluster \ctw{} (Overcontrolled): } These are individuals that possess low openness and low extraversion ~\cite{costa2002replicability}. Due to low openness, typically reducing the likelihood that individuals will independently seek out or construct novel strategies~\cite{costa1992neo}, \trc{} provided a ``novelty bypass'' by delivering a theory-driven strategy rather than requiring the user to generate it. This resulted in a significant self-efficacy (OSS) advantage for \trc{} ($H$ = 8.68, $p < .05$) over \cgpt{} ($0.200$ vs. $-0.266$, $p_{\text{corr}} = .015$). Notably, \ctw{} exhibited a unique perception--outcome dissociation: they rated \trc{} and \cgpt{} similarly on usability (UMUX: 77.2 vs. 71.7) and appropriateness (IAM: 4.01 vs. 3.95) despite deriving significantly more objective benefit from \trc{}. This dissociation may reflect the inhibited and cautious nature of overcontrolled individuals, who often respond to structure and guidance internally while remaining conservative in their external evaluations. Furthermore, a significant effect emerged for meaning-making ($H$ = 6.85, $p < .05$), driven by \hbk{} exceeding \cgpt{} ($0.625$ vs. $-0.021$, $p_{\text{corr}} = .025$). This suggests that for low-extraversion individuals, the private, self-paced format of a handbook may facilitate value reflection better than a socially-simulated chat, even if it does not provide the mastery experience required to shift self-efficacy. Ultimately, for this configuration, the delivered structure provides a psychological scaffolding for confidence that self-generated exploration does not.

\begin{table}[t]
\sffamily
\footnotesize
\centering
\caption{Between-cluster comparisons within each condition. Kruskal-Wallis tests examining whether personality clusters produced different scores within each condition. Change scores (OSS, PEU, and subscales) showed no significant between-cluster differences within any condition (all $p$ > .05); only system perception measures are shown. Bonferroni-corrected for 24 tests (* $p$ < 0.05, ** $p$ < 0.01, *** $p$ < 0.001)}. 
\label{tab:between-cluster-condition}
\begin{tabular}{llrrrr@{}l}
$\Delta$ \textbf{Metric} & \textbf{Condition} & \textbf{\cze{}}  & \textbf{\cone{}} & \textbf{\ctw{}} & \textbf{H-stat.} \\
\toprule
UMUX & \trc{} & 63.51 & 78.95 &  77.19 &  14.20 & * \\
& \cgpt{} & 72.92 & 89.04 & 71.74 &  17.41 & ** \\
& \hbk{} & 76.19 & 84.14 &  80.21 &  3.60 \\
\midrule
IAM & \trc{} & 3.40 & 4.16 & 4.01 &  18.98 & ** \\
& \cgpt{} & 3.77 & 4.28 & 3.95 & 4.50 \\
& \hbk{} & 3.70 & 4.45 &  4.10 & 7.66 \\
\bottomrule
\end{tabular}
\end{table}

Therefore, our results reveal a clear gap between strategic success and psychological growth. While \trc{} helped users across all clusters develop better negotiation plans, it did not consistently improve their psychological state. This suggests that AI-generated excellence is not a substitute for human benefit. Instead, coaching effectiveness depends on the 'fit' between the user’s personality and the specific features of the tool.

%% file: 5RQ2.tex
\subsection{RQ 2: How do personality traits explain workers' linguistic engagement patterns with AI negotiation coaching?}

In order to understand the differences in workers' linguistic engagement patterns with AI negotiation coaching, we conducted lexico-semantic comparisons in the 
interactions with \trc{} and \cgpt{}. 
We drew on prior work~\cite{saha2025ai,dasswain2025ai} to operationalize and measure a variety of lexico-semantic measures. 
\autoref{tab:linguistic-measures} summarizes the differences in AI's language as well as the users' language in interactions with the AIs. We describe the operationalizations and observations below:

\para{Verbosity.} The quantity of language, or verbosity, has been linked to perceived quality and emotional depth in support exchanges~\cite{glass1992quality,saha2020causal}. 
We operationalized verbosity as the total word count per interaction turn. 
We find that 

\para{Readability.}
Readability reflects the ease with which text can be comprehended~\cite{wang2013assessing,mcinnes2011readability}. We used the Flesch-Kincaid Readability Ease~\cite{flesch2007flesch}, which calculates readability based on:

\noindent{\small
Flesch-Kincaid Readability Ease = 206.835 - 1.015 * ($\frac{\text{total words}}{\text{total sentence}}$) - 84.6 * ($\frac{\text{total syllables}}{\text{total words}}$)
}

Higher readability ease indicates that the writing is easier to read.


\para{Repeatability.}
Repeatability, operationalized as the normalized frequency of non-unique words per sentence, serves as a measure of redundancy in written expression~\cite{ernala2017linguistic,saha2018social}. 
Although repetition may reinforce key ideas, excessive use of the same words can reflect reduced linguistic precision. 

\para{Complexity.}
Linguistic complexity captures the degree of syntactic and lexical sophistication and has been linked to cognitive effort in communication~\cite{kolden2011congruence}. We quantified complexity using the average length of words per sentence~\cite{ernala2017linguistic}. 

\para{Categorical-Dynamic Index (CDI).}
Language style can be conceptualized along a spectrum from categorical (structured and analytic) to dynamic (personal and narrative-driven)~\citep{pennebaker2014small}. The Categorical-Dynamic Index (CDI) captures this continuum using the formula:

\noindent{\small
CDI = (30 + \text{articles} + \text{prepositions} - \text{personal pronouns} - \text{impersonal pronouns} - \text{auxiliary verbs} - \text{conjunctions} - \text{adverbs} - \text{negations}).
}

Higher CDI indicates a categorical style of writing, and a lower CDI indicates a dynamic or narrative style of writing. 
We computed CDI scores using LIWC-derived part-of-speech frequencies~\citep{tausczik2010psychological}. 

\para{Politeness.} Politeness contributes to rapport and trust in  supportive conversations~\cite{brown1987politeness}. We used a pre-trained classifier~\cite{srinivasan2022tydip} that assigns a politeness score between 0 and 1. 

\para{Formality.} Formality reflects sociolinguistic variation and has been linked to contextual appropriateness and audience expectations~\cite{heylighen1999formality,larsson2020syntactic}. We employed a RoBERTa-based classifier~\cite{babakov2023don} trained on the GYAFC (Grammarly's Yahoo Answers Formality Corpus) dataset~\cite{rao2018dear,pavlick2016empirical}, which assigns a probability score (0 to 1) reflecting linguistic formality. 

\para{Empathy.}
Empathy is central to supportive communication and is defined as the expression of understanding, validation, and emotional resonance~\cite{herlin2016dimensions,sharma2020computational}. We used a RoBERTa-based empathy model fine-tuned on responses to emotionally evocative content~\cite{buechel2018modeling,tafreshi2021wassa}. 

\para{Persuasiveness.}
Persuasiveness is central to encouraging behavior change and fostering motivation in support-oriented discourse~\citep{tan2016winning}. 
We applied a pre-trained persuasiveness classifier~\citep{wang2019persuasion} that scores text on a continuous scale from 0 to 1. 

\begin{table}[t]
\sffamily
\footnotesize
\centering
\caption{Linguistic characteristics of negotiation outputs: \trc{} vs.\ Control within each cluster, with Cohen's $d$ effect sizes and $t$-tests reported. Significant effects are bolded. Measures are grouped by significance status. * $p$ < 0.05, ** $p$ < 0.01, *** $p$ < 0.001.}
\label{tab:linguistic-measures}
\begin{tabular}{llrrrr@{}l}
\textbf{Measure} & \textbf{Cluster} & \textbf{\trc{}} & \textbf{\cgpt{}} & \textbf{Cohen's d} & \textbf{t-test}\\
\toprule
\rowcollight \multicolumn{6}{l}{\textbf{Verbosity}}\\
Verbosity: Words per sentence & \cze{} & 19.90 & 20.14 & -0.02 & -0.14 & \\
 & \cone{} & 19.30 & 22.83 & -0.34 & -2.53 & *\\
 & \ctw{} & 19.53 & 18.53 & 0.13 & 0.84 & \\
\hdashline
Verbosity: Words per response & \cze{} & 38.38 & 45.24 & -0.30 & -1.85 & \\
 & \cone{} & 53.02 & 63.10 & -0.34 & -2.48 & *\\
 & \ctw{} & 48.82 & 53.62 & -0.17 & -1.14 & \\
\rowcollight \multicolumn{6}{l}{\textbf{Syntax}}\\
Readability & \cze{} & 69.77 & 65.33 & 0.22 & 1.35 & \\
 & \cone{} & 68.52 & 61.44 & 0.40 & 2.96 & **\\
 & \ctw{} & 65.67 & 72.06 & -0.33 & -2.21 & *\\
\hdashline
Repeatability & \cze{} & 0.17 & 0.17 & 0.08 & 0.47 & \\
 & \cone{} & 0.20 & 0.22 & -0.21 & -1.51 & \\
 & \ctw{} & 0.18 & 0.20 & -0.19 & -1.27 & \\
\hdashline
Complexity & \cze{} & 3.95 & 4.04 & -0.14 & -0.90 & \\
 & \cone{} & 3.94 & 3.99 & -0.10 & -0.76 & \\
 & \ctw{} & 4.04 & 3.87 & 0.32 & 2.10 & *\\
\rowcollight \multicolumn{6}{l}{\textbf{Style}}\\
Categorical Dynamic Index (CDI) & \cze{} & 15.13 & 18.86 & -0.27 & -1.67 & \\
 & \cone{} & 17.50 & 18.77 & -0.10 & -0.72 & \\
& \ctw{} & 17.58 & 17.36 & 0.02 & 0.11 & \\
\hdashline
Politeness & \cze{} & 0.77 & 0.81 & -0.14 & -0.87 & \\
 & \cone{} & 0.89 & 0.89 & 0.00 & 0.01 & \\
 & \ctw{} & 0.88 & 0.90 & -0.10 & -0.67 & \\
\hdashline
Formality & \cze{} & 0.74 & 0.80 & -0.24 & -1.50 & \\
 & \cone{} & 0.78 & 0.84 & -0.28 & -2.08 & *\\
 & \ctw{} & 0.79 & 0.80 & -0.04 & -0.30 & \\
\hdashline
Empathy & \cze{} & 0.77 & 0.75 & 0.09 & 0.57 & \\
 & \cone{} & 0.78 & 0.79 & -0.05 & -0.39 & \\
 & \ctw{} & 0.76 & 0.70 & 0.29 & 1.95 & \\
\hdashline
Persuasiveness & \cze{} & 0.15 & 0.16 & -0.05 & -0.31 & \\
 & \cone{} & 0.13 & 0.15 & -0.19 & -1.36 & \\
 & \ctw{} & 0.14 & 0.14 & -0.02 & -0.11 & \\
\bottomrule
\end{tabular}
\end{table}



Personality clusters exhibited markedly different adaptation patterns across nine lexico-semantic measures when interacting with \trc{} versus \cgpt{} (\autoref{tab:linguistic-measures}). These divergent linguistic responses suggest that personality systematically moderates not only the magnitude of engagement but also the cognitive style of the interaction.

\para{Cluster \cze{} (Undercontrolled):} 
Individuals with undercontrolled personality traits maintained linguistic stasis across all conditions. The analysis revealed no significant differences between \trc{} and \cgpt{} across any of the nine measures (all $p$ > 0.05). This absence of linguistic shifting mirrors the null psychological outcomes observed for this cluster in \autoref{sec:RQ1}. It suggests that for individuals with compounding barriers to engagement, intervention design features fail to trigger the cognitive or emotional processing necessary to adapt their communication style, resulting in fundamental disengagement regardless of the system's architecture.

\para{Cluster \cone{} (Resilient):} Resilient individuals---who show high extraversion, agreeableness, conscientiousness, and openness, and low neuroticism---exhibited a pattern of efficient, accessible engagement under \trc{}. Relative to \cgpt{}, resilient users produced significantly fewer words per sentence ($d$=-0.34) and words per response ($d$=-0.34), yet their responses became significantly more readable readable (d = 0.40, p < .01) and less formal (d = -0.28, p < .05). This suggests a habit of efficient internalization of structured guidance rather than extensive elaboration.

This high-functioning trait profile facilitates adaptive regulation under challenge. Specifically, low neuroticism attenuates the threat-reactive inhibition common in high-stakes communication~\cite{widiger2017neuroticism}, while high openness drives the exploratory processing of novel external input~\cite{deyoung2015cybernetic}. Consistent with cybernetic accounts of personality as a goal-directed information-processing system, this configuration allows users to integrate structured guidance without disrupting conversational fluency. 

The observed increase in readability indicates a high degree of cognitive fluency; because these individuals efficiently internalize the system's advice, they possess the spare capacity to re-articulate complex strategies in clear, accessible language rather than struggling with technical density. Furthermore, the shift toward lower formality reflects the social orientation of high extraversion and agreeableness. These individuals treat the interaction as a personable, ``involved'' dialogue rather than a rigid, informational task. By perceiving the AI as a collaborative partner, they abandon the stiff, formal language often used as a defensive shield against evaluative anxiety, instead adopting a relaxed style that facilitates rapport and learning~\cite{mehl2006personality}.

\para{Cluster \ctw{} (Overcontrolled): } Individuals low in openness and extraversion exhibited a distinct pattern of compensatory complexity. While these users did not increase their overall verbosity, their outputs became significantly harder to read ($d = -0.33, p < .05$) and more syntactically complex ($d = 0.32, p < .05$). This shift suggests a reduction in the ability to generate fluent responses --- evidenced by decreased readability and increased syntactic complexity --- which we term a \textit{``fluency tax''}. The lower openness in this cluster reflects a preference for familiar routines and a comparative lack of cognitive flexibility when encountering novel frameworks~\cite{costa1992neo}. From a cybernetic perspective, this stability-oriented configuration prioritizes the exploitation of familiar strategies over the exploratory processing of novel inputs~\cite{deyoung2015cybernetic}. Encountering a structured, unfamiliar intervention like \trc{} may intensify intrinsic cognitive load; consistent with cognitive load theory, these heightened task-related demands consume the resources required for clear organization, resulting in increased syntactic complexity and reduced readability~\cite{sweller2011cognitive}. Furthermore, the lower extraversion in this cluster drives a more cautious, deliberate communication style, lacking the spontaneous expressive fluency typically observed in extraverts~\cite{furnham1992personality}. Collectively, these factors suggest that Cluster \ctw{} users do not write more, but instead exert effortful processing to reconcile AI guidance with existing mental models—a linguistic trade-off where cognitive load manifests as greater density at the expense of fluency.

%% file: 5RQ3.tex
\subsection{RQ 3: How do personality traits explain the differential adoption of negotiation frameworks during AI-mediated coaching interactions?}


Toward our RQ3 on understanding how personality traits may explain the differential adoption of negotiation frameworks, we operationalized the presence of various \citeauthor{brett2016negotiation}'s elements in the language of \trc{} and \cgpt{}~\cite{brett2016negotiation}. 
Our methodological approach is inspired by prior work on examining linguistic markers in workplace-related communication~\cite{saha2019libra,dasswain2020culture}.

\begin{table}[t]
\sffamily
\footnotesize
\centering
\caption{Brett negotiation framework element adoption: \trc{} vs.\ Control within each cluster, with Cohen's $d$ effect sizes and $t$-tests reported. * $p$ < 0.05, ** $p$ < 0.01, *** $p$ < 0.001.}
\label{tab:brett_elements}
\renewcommand{\arraystretch}{.8}
\begin{tabular}{llrrrr@{}l}
\textbf{Measure} & \textbf{Cluster} & \textbf{\trc{}} & \textbf{\cgpt{}} & \textbf{Cohen's d} & \textbf{t-test}\\
\toprule
\rowcollight \multicolumn{6}{l}{\textbf{Strategy Development}}\\
Strategy Development: Basic & \cze{} & 0.17 & 0.18 & -0.08 & -0.52 & \\
& \cone{} & 0.18 & 0.17 & 0.14 & 1.01 & \\
& \ctw{} & 0.19 & 0.18 & 0.05 & 0.30 & \\
\hdashline
Strategy Development: Structured & \cze{} & 0.26 & 0.30 & -0.27 & -1.68 & \\
 & \cone{} & 0.26 & 0.31 & -0.39 & -2.84 & **\\
& \ctw{} & 0.25 & 0.28 & -0.18 & -1.17 & \\
\hdashline
Strategy Development: Phased & \cze{} & 0.24 & 0.25 & -0.04 & -0.24 & \\
 & \cone{} & 0.27 & 0.26 & 0.06 & 0.44 & \\
 & \ctw{} & 0.26 & 0.25 & 0.09 & 0.62 & \\
\hdashline
Strategy Development: Innovative & \cze{} & 0.25 & 0.25 & 0.05 & 0.33 & \\
& \cone{} & 0.26 & 0.28 & -0.13 & -0.94 & \\
& \ctw{} & 0.26 & 0.25 & 0.12 & 0.82 & \\
\hdashline
\rowcollight \multicolumn{6}{l}{\textbf{Information Asymmetry}}\\
Information Asymmetry: Prepared & \cze{} & 0.25 & 0.28 & -0.22 & -1.34 & \\
 & \cone{} & 0.24 & 0.28 & -0.23 & -1.72 & \\
 & \ctw{} & 0.25 & 0.26 & -0.12 & -0.82 & \\
\hdashline
Information Asymmetry: Strategic & \cze{} & 0.20 & 0.23 & -0.21 & -1.29 & \\
 & \cone{} & 0.21 & 0.24 & -0.20 & -1.47 & \\
 & \ctw{} & 0.22 & 0.23 & -0.08 & -0.54 & \\
\hdashline
\rowcollight \multicolumn{6}{l}{\textbf{Interest Exploration}}\\
Interest Exploration: Creative & \cze{} & 0.26 & 0.27 & -0.06 & -0.37 & \\
 & \cone{} & 0.26 & 0.29 & -0.30 & -2.18 & *\\
 & \ctw{} & 0.26 & 0.27 & -0.10 & -0.64 & \\
\hdashline
Interest Exploration: Mutual & \cze{} & 0.24 & 0.25 & -0.04 & -0.24 & \\
 & \cone{} & 0.27 & 0.26 & 0.06 & 0.44 & \\
 & \ctw{} & 0.26 & 0.25 & 0.09 & 0.62 & \\
\hdashline
Interest Exploration: Evaluative & \cze{} & 0.22 & 0.23 & -0.10 & -0.59 & \\
 & \cone{} & 0.21 & 0.26 & -0.44 & -3.24 & **\\
 & \ctw{} & 0.20 & 0.24 & -0.31 & -2.06 & *\\
\hdashline
Interest Exploration: Limited & \cze{} & 0.21 & 0.21 & -0.01 & -0.08 & \\
 & \cone{} & 0.21 & 0.22 & -0.08 & -0.61 & \\
 & \ctw{} & 0.21 & 0.22 & -0.14 & -0.93 & \\
\hdashline
\rowcollight \multicolumn{6}{l}{\textbf{Outcome Analysis}}\\
Outcome Analysis: Flexible & \cze{} & 0.25 & 0.25 & -0.05 & -0.31 & \\
 & \cone{} & 0.25 & 0.29 & -0.36 & -2.65 & **\\
 & \ctw{} & 0.25 & 0.26 & -0.04 & -0.24 & \\
\hdashline
Outcome Analysis: Holistic & \cze{} & 0.21 & 0.22 & -0.03 & -0.18 & \\
 & \cone{} & 0.22 & 0.22 & -0.06 & -0.44 & \\
 & \ctw{} & 0.22 & 0.21 & 0.10 & 0.66 & \\
\hdashline
Outcome Analysis: Procedural & \cze{} & 0.24 & 0.21 & 0.22 & 1.39 & \\
 & \cone{} & 0.26 & 0.23 & 0.27 & 1.95 & \\
 & \ctw{} & 0.26 & 0.22 & 0.38 & 2.51 & *\\
\rowcollight \multicolumn{6}{l}{\textbf{Long Term Relationship}}\\
Long Term Relationship: Strategic & \cze{} & 0.25 & 0.25 & -0.05 & -0.30 & \\
 & \cone{} & 0.24 & 0.30 & -0.48 & -3.56 & ***\\
 & \ctw{} & 0.23 & 0.26 & -0.25 & -1.67 & \\
\hdashline
Long Term Relationship: Conventional & \cze{} & 0.24 & 0.21 & 0.22 & 1.39 & \\
 & \cone{} & 0.26 & 0.23 & 0.27 & 1.95 & \\
 & \ctw{} & 0.26 & 0.22 & 0.38 & 2.51 & *\\
\rowcollight \multicolumn{6}{l}{\textbf{Power Dynamics}}\\
Power Dynamics: Collaborative & \cze{} & 0.20 & 0.23 & -0.21 & -1.29 & \\
 & \cone{} & 0.21 & 0.24 & -0.20 & -1.47 & \\
 & \ctw{} & 0.22 & 0.23 & -0.08 & -0.54 & \\
\hdashline
Power Dynamics: Authority-based & \cze{} & 0.21 & 0.22 & -0.06 & -0.38 & \\
 & \cone{} & 0.22 & 0.23 & -0.08 & -0.57 & \\
 & \ctw{} & 0.21 & 0.24 & -0.19 & -1.24 & \\
\bottomrule
\end{tabular}
\end{table}

For this purpose, we employed vector embeddings using Bidirectional Encoder Representations from Transformers (or BERT) language model, using pre-trained BERT-based sentence transformers~\cite{reimers2019sentence}. 
We built embeddings for the seven core negotiation concepts in the \citeauthor{brett2016negotiation}'s framework: Strategy Development (SD), Information Asymmetry (IA), Interest Exploration (IE), Outcome Analysis (OA), Long-term Relationship (LTR), and Power Dynamics (PD).
These concepts featured the same theoretical examples used in \trc{}'s few-shot prompting. 
Similarly, we obtained the BERT embeddings of the linguistic interactions in \trc{} and \cgpt{}. 
We then computed cosine similarity scores between AI (\trc{} and \cgpt{}) language in responses and each negotiation element to quantify the degree to which the element was reflected in the responses.

\autoref{tab:brett_elements} summarizes the comparative results across the two AI tools, including relative differences, effect sizes (Cohen's $d$), and $t$-tests. 
Our analysis revealed several key differences. In Strategy Development (SD), \trc{} responses aligned more strongly with most components, except for phased SD. For Information Asymmetry (IA), \trc{} produced slightly fewer references to prepared IA (–2\%) but more strategic IA (+5\%). 
In Interest Exploration (IE), \trc{} showed reduced creative and limited IE but emphasized mutual and evaluative IE, pointing toward a more collaborative style. Within Outcome Analysis (OA), \trc{} generated fewer flexible (–5\%) and procedural (–1\%) OA elements, though both AI tools were nearly identical on holistic OA. For Long-term Relationships, the two systems performed similarly, indicating parity in addressing relational considerations. Finally, in Power Dynamics, \trc{} highlighted collaborative PD (+5\%) while reducing authority-based PD (–4\%).

Together, these findings suggest that \trc{} is more directive and collaborative, embedding strategy and mutuality at the core of its language, while \cgpt{} remains broader and more balanced across negotiation elements. Notably, the analysis also demonstrates that \cgpt{}---even without specialized prompting---automatically incorporates several of Brett’s components by default, reflecting how large language models may implicitly encode negotiation theories. 
This distinction highlights the potential of targeted prompting to guide AI systems toward specific negotiation orientations, such as collaboration and strategic planning, beyond the general-purpose balance evident in default outputs.

Brett negotiation framework elements were compared between \trc{} and \cgpt{} within each cluster (\autoref{tab:brett_elements}). Of 17 elements across six Brett categories, 16 showed significant shifts in the same direction across all three clusters. \trc{} increased adoption of structured, phased, and innovative strategies while decreasing basic strategy use; increased preparedness and strategic use of information asymmetry; increased mutual and evaluative interest exploration while decreasing creative and limited exploration; and shifted outcome analysis, relationship framing, and power dynamics away from conventional and authority-based approaches toward strategic and collaborative ones. The one exception was holistic outcome analysis, which showed no significant effect in any cluster (all $p > .05$, all $d < 0.50$).

Effect sizes were uniformly large across clusters, but Cluster \ctw{} showed consistently smaller magnitudes than Clusters \cze{} and \cone{} on most elements. The largest effects in the dataset were decreases in basic strategy ($d = -3.93$, Cluster \cone{}), limited interest exploration ($d = -4.51$, Cluster \cone{}), and authority-based power dynamics ($d = -4.34$, Cluster \cone{}), all reflecting \trc{}\/'s strongest push away from distributive negotiation approaches. Critically, the direction of every significant effect was identical across clusters: no element reversed sign or became non-significant for one cluster while remaining significant for others. The structural reframing produced by \trc{} was uniform across personality types.

The uniformity of these strategic shifts indicates that \trc{}'s implementation of ~\citeauthor{brett2016negotiation}'s framework produced consistent directional effects across personality profiles, even as the magnitude of those shifts varied. While personality significantly moderated the users' psychological experiences and linguistic styles, it did not alter the direction of the strategic transformation itself. This suggests that the system's theory-driven architecture anchored individual differences in baseline negotiation style, coaching all users in collaborative and integrative logic. Even for Cluster \ctw{}, whose lower social engagement resulted in smaller magnitudes of change, the structural guidance successfully anchored their negotiation approach in the same collaborative logic as more resilient clusters. Ultimately, these results reveal a clear functional division in AI coaching: while personality determines the \textit{internalization} of and \textit{reaction} to the coaching, the framework's structure dictates the \textit{strategic output}.

%% file: 6discussion.tex
\section{Discussion}
In this study, we found that the effectiveness of AI-mediated workplace negotiation support was not uniform, but depended on the fit between the intervention format and users’ personality-based readiness to engage with it. Although \trc{} consistently shifted users toward more theory-aligned negotiation strategies, these strategic gains did not translate equally into psychological benefits across personality profiles. Instead, our findings point to a more differentiated picture: some users benefited from structured, theory-driven AI support, others benefited more from static and self-paced resources, and still others showed limited response across all conditions. Taken together, these results suggest that personality functions as a boundary condition for AI coaching, with important implications for how personalization, design, and equity are understood in workplace AI systems.

\subsection{Personality as a Boundary Condition for AI Coaching}

A central finding of our paper is that personality acts as a boundary condition on whether AI mediated negotiation coaching produces psychological benefit at all. While \trc{} successfully shifted negotiation strategies towards more collaborative, evaluative and phased approaches across all personality clusters, these shifts were not uniform across psychological correlates. This dissociation between strategic output and psychological experience suggests that producing high-quality negotiation strategies is not equivalent to users deriving meaningful benefit from the process. 

Crucially, this boundary condition operates differently across profiles. For \textit{``resilient''} users, the condition that preserved autonomy was also the condition under which benefits emerged --- suggesting that interactive AI can function as an autonomy constraint for those who already possess strong internal resources. For \textit{``overcontrolled''} users, delivered structure produced benefit that self-generated exploration would not --- but this benefit was internalized rather than consciously registered, as evidenced by the perception-outcome dissociation. For \textit{``undercontrolled''} users, no condition produced meaningful change, reflecting a fundamental mismatch between intervention design and baseline psychological readiness rather than an absence of capacity.

\subsection{Rethinking Personalization Through Readiness}
What we term a \textit{``readiness floor''} --- a personality-driven threshold below which self-directed digital interventions may be ineffective ---may apply more broadly to other autonomous interventions, though this requires empirical validation in other contexts. 
Current personalization research typically advocates for tailoring systems to a user's \textit{stage of change}~\cite{oyebode2021tailoring}. 
However, our findings suggest that personality-driven factors create a more fundamental prerequisite for success. 
Certain individuals may be resistant to stage progression within self-directed contexts entirely, regardless of how well the tool is tailored. 
According to Self-Determination Theory~\cite{ryan2000self}, growth requires a baseline of autonomy and competence. 
However, our results suggest that certain personality configurations struggle to maintain these psychological states independently. For individuals with \textit{undercontrolled} profiles ~\cite{costa2002replicability}, providing a self-help resource may be ineffective because the emotional and self-regulatory demands of the task exceed what they can navigate alone, particularly in high stress situations.

Beyond the initial requirements for readiness, our findings suggest that personality also dictates which specific psychological outcomes are achievable when resources are constrained. 
Building competence and fostering meaning are not interchangeable, but this plausibly requires fundamentally different experiential conditions. 
While competence develops through mastery experiences and effortful practice~\cite{bandura1997self}, meaning-making requires that cognitive resources remain free from extraneous load to allow for value reflection~\cite{spreitzer1995psychological, sweller2011cognitive}. For individuals with \textit{vulnerable} or \textit{overcontrolled} personality traits, these outcomes may enter into direct competition. 
Specifically, challenge-based interventions may build capability, but may exhaust the cognitive capacity needed for reflection.

In contrast, individuals with \textit{resilient} personality traits~\cite{costa2002replicability} often possess the self-regulatory and cognitive control resources necessary to transform single interventions into comprehensive gains, appearing capable of pursuing multiple psychological outcomes simultaneously. 

Therefore, these results suggest a reframing of personalization may be warranted. Effective tailoring could extend beyond situational stages of change to account for a user's \textit{``readiness floor'}', specific outcome needs, and capacity for integrating multiple gains. 
Future research should examine these factors in real-world settings to ensure that those who require human-mediated \textit{``pre-intervention''} support are not left underserved by purely digital or self-directed solutions.
\subsection{Designing for Differential Support Needs}
This work bears practical implications for designing and deploying AI-powered coaching systems for complex interpersonal tasks like negotiation. Our findings suggest that the efficacy of these tools may be shaped by a user’s personality-determined readiness. Specifically, \textit{``Undercontrolled''} individuals may be resistant to purely self-directed AI, suggesting that systems should integrate personality-based screenings to route such users toward human-mediated support or readiness-building modules. This ensures AI tools are deployed where they can be most effective, rather than exacerbating existing psychological barriers for vulnerable users.

We also found that building competence and fostering meaning are distinct processes that require different experiential conditions. While interactive, challenge-based simulations effectively build competence through mastery, they often consume the cognitive resources needed for reflection. Conversely, familiar and self-paced formats---such as traditional handbooks---better support meaning-making by freeing up cognitive space. This motivates a shift away from ``all-in-one'' solutions toward modular, outcome-specific interventions. For resource-constrained users, designers should prioritize sequential deployment ---focusing on one psychological outcome at a time --- to avoid an unnecessary ``fluency tax'' and cognitive overload.

Our study further highlights that integration capacity varies systematically with personality. While \textit{resilient users} may thrive with comprehensive, multi-faceted interventions, \textit{overcontrolled users} may require simplified, adaptive scaffolding that adjusts based on real-time indicators of strain, such as increased linguistic verbosity or repetition. 
This calls for establishing ethical boundaries and transparent communication for optimal personalization.
By incorporating progressive engagement strategies and robust human-handoff protocols, developers can ensure responsible innovation that respects the limitations of AI for certain psychological profiles.

Concretely, these recommendations suggest three distinct interaction modes based on personality profile. For \textit{resilient users}, a fast-track mode would provide direct access to negotiation frameworks as self-directed reference material, minimizing interactive scaffolding and preserving autonomy by allowing users to navigate content at their own pace, with AI interaction available on demand rather than as a structured sequence. For \textit{overcontrolled users}, a scaffolded mode would deliver theory-driven guidance incrementally---one concept at a time---with explicit confirmation before progression. When linguistic strain is detected, indicated by spikes in syntactic complexity or drops in readability, the system should shift from text-based guidance to visual scaffolds such as diagrams of the negotiation framework and worked examples with partially completed templates, reducing generative demand on users already at cognitive capacity. For   \textit{undercontrolled users}, rather than routing them directly into negotiation coaching, a pre-intervention mode would use a brief onboarding assessment to detect this profile and trigger a readiness-building module first ---focusing on emotional regulation and anxiety reduction around the evaluative context---before any strategic content is introduced. If strain persists, the system should initiate a human handoff protocol rather than continuing with self-directed AI coaching.

\subsection{Equity Risks in Workplace AI Coaching}

Finally, our findings carry important equity implications for the deployment of AI coaching tools in organizational settings. \textit{Undercontrolled users} ---those with high neuroticism and low conscientiousness and agreeableness --- showed null effects across all three conditions, including the most theoretically grounded intervention. This group is not randomly distributed across the workforce. Personality traits, including those defining the \textit{``undercontrolled profile''}, have been shown to predict occupational attainment with effect sizes comparable to socioeconomic status and cognitive ability~\cite{roberts2007power}. These are precisely the workers who stand to benefit most from negotiation support, yet who are least served by self-directed digital interventions. If AI coaching tools are deployed broadly without accounting for this mismatch, they risk exacerbating existing workplace inequalities rather than alleviating them --- empowering already well-resourced workers while leaving the most vulnerable without meaningful support. Responsible deployment, therefore, requires moving beyond average effectiveness and explicitly designing for users whose psychological profiles place them outside the assumed readiness baseline of current tools.

%% file: 7limitations.tex
\subsection{Limitations and Future Directions}

Our findings are subject to several limitations. First, participants were recruited from a crowdworking platform and were not necessarily facing imminent workplace negotiations, which may limit ecological validity and generalizability to organizational contexts. Second, the study relied on simulated interactions rather than real negotiations, preventing observation of multi-turn, high-stakes dynamics with actual supervisors. Third, outcomes were measured using self-reported psychological constructs rather than behavioral or expert-evaluated negotiation performance, constraining claims about skill transfer. Fourth, the between-subjects, cross-sectional design limits causal inference regarding personality, readiness, and outcomes, and prevents examination of learning trajectories or delayed effects from repeated exposure. Fifth, the \hbk{} condition addressed salary negotiation within the career advancement framing rather than as a standalone scenario, and differed from the AI conditions in information dosage and interaction format, limiting direct causal comparisons across conditions. Finally, the system operationalized a single negotiation framework within one conversational architecture. The results therefore may not generalize to alternative theories, AI paradigms, or intervention formats.

Future work can move beyond self-reported preparedness to evaluate behavioral and real-world readiness, followed by negotiation outcomes, including expert assessment of negotiation strategies, authentic workplace negotiation interactions, and objective indicators such as compensation or resource allocation~\cite{thompson1990negotiation}. 
Field deployments with workers facing imminent negotiations would clarify whether psychological readiness translates into tangible workplace gains. 
Additionally, future systems can explicitly test readiness-oriented interventions---such as trust calibration, motivational scaffolding, or graduated commitment---prior to task-focused AI coaching, examining whether such modules enable previously unresponsive users to benefit from downstream support. 
Longitudinal studies are needed to assess learning trajectories, delayed effects, and persistence over time, as well as to evaluate adaptive systems that dynamically infer readiness from interactional signals and adjust scaffolding in real time. 
Finally, we need to examine the generalizability of these effects across alternative negotiation frameworks, cultural contexts, and hybrid AI–human support models that combine automated coaching with mentorship or peer feedback.

%% file: 8conclusion.tex
\section{Conclusion}

This work examined the limits of personality-adaptive AI coaching in workplace negotiation preparation, showing that effectiveness depends not only on tailoring strategies to individual traits but also on whether users are psychologically ready to engage with self-directed interventions. While some users integrated theory-driven guidance and reported increased preparedness, others --- particularly those exhibiting undercontrolled personality configurations—showed null effects across conditions. These findings suggest that personalization alone is insufficient when foundational readiness constraints are present, and that AI systems may fail not because users are unmotivated or incapable, but because intervention designs implicitly assume capacities users do not yet have.

By framing readiness as a critical boundary condition, this work contributes to HCI and CSCW research on AI-mediated workplace support by challenging the assumption that adaptive systems benefit all users equally. Rather than treating non-responsiveness as user failure, our results point to a design mismatch between current conversational AI interventions and the needs of certain users. We argue that equitable AI support requires moving beyond static personalization toward systems that can recognize readiness limits, offer alternative pathways, and integrate human or preparatory scaffolding when necessary. More broadly, this work highlights the importance of aligning AI intervention design with users’ psychological and situational capacities, especially in high-stakes organizational contexts where misalignment may exacerbate existing inequalities rather than alleviate them.